\documentclass[10pt,aps,prd]{revtex4-1} 
\usepackage{amsfonts,amsmath,amssymb,mathptmx,amsthm,bm}
\usepackage{array,mathtools,latexsym,mathrsfs,soul,slashed}
\usepackage{graphicx,graphics,subfigure}
\usepackage{ctable,tabularx,multirow}
\usepackage[colorlinks=true,linkcolor=blue,citecolor=blue,urlcolor=blue]{hyperref}

\DeclareSymbolFont{symbols} {OMS}{cmsy}{m}{n}

\def\be{\begin{equation}}
\def\ee{\end{equation}}
\def\bea{\begin{eqnarray}}
\def\eea{\end{eqnarray}}
\def\ba{\begin{aligned}}
\def\ea{\end{aligned}}
\def\nn{\nonumber}
\def\p{\partial}
\def\cN{\mathcal{N}}
\def\cV{\mathcal{V}}

\begin{document}


\title{Thermodynamics and topological classifications of static non-extremal four-charge AdS
black hole in the five-dimensional $\cN = 2$, $STU-W^2U$ gauged supergravity}

\author{Di Wu}
\email{wdcwnu@163.com}

\author{Shuang-Qing Wu}
\email{sqwu@cwnu.edu.cn}

\affiliation{School of Physics and Astronomy, China West Normal University,
Nanchong, Sichuan 637002, People's Republic of China}

\date{\today}

\begin{abstract}
In this paper, we investigate thermodynamical aspects of a novel static non-extremal black hole
solution in the five-dimensional anti-de Sitter (AdS) space carrying four independent electric
charge parameters and with a negative cosmological constant, within the framework of $D=5$, $\cN
= 2$ gauged supergravity coupled to three vector multiplets specified by the pre-potential $\cV
= STU -W^2U \equiv 1$. A central outcome of our work is the demonstration that the thermodynamics
of this solution consistently satisfies both the differential and integral forms of the mass formulae
of black hole thermodynamics. When the fourth charge is set to zero, our configuration reduces to
the known three-charge static non-extremal AdS$_5$ black hole within the gauged $STU$ model, and its
thermodynamic behavior connects smoothly to that limit. Finally, we also provide a brief discussion
about topological classifications of its thermodynamics.
\end{abstract}

\maketitle


\section{Introduction}

With the advent of the AdS/CFT correspondence, it has become of particular interest to study the AdS
black hole solutions of gauged supergravities in five and other dimensions, since they can provide
useful test-beds for this conjectured AdS/CFT correspondence. Although there has been much progress
in constructing black hole solutions of gauged supergravity during the past years, there remains to
be of great difficulty in the construction of charged rotating non-extremal black holes in gauged
supergravity theory (see, a brief summary in Ref. \cite{CQG27-205009}), for example, an exact rotating
charged non-extremal AdS$_6$ black hole solution with two independent rotations and two different
electric charges in the six-dimensional (probably unclear) gauged supergravity theory still remains
elusive.

For a variety of reasons, significant attention has recently been focused on studying black holes in
high-dimensional supergravity, which exhibit distinctive features absent in their four-dimensional
counterparts. Among various supergravity theories, the five-dimensional $\cN = 2$ supergravity stands
out as a particularly rich and tractable framework. Its structure allows for a systematic construction
of black hole solutions while incorporating supersymmetry, making it an ideal setting for exploring
the interplay between black hole physics and fundamental symmetries. The bosonic sector of the
five-dimensional $\cN = 2$ supergravity includes the gravity multiplet and $n$ Abelian vector
multiplets. The interactions are governed by a symmetric tensor $C_{IJK}$, while supersymmetry
restricts the scalar manifolds to ``very special geometry" \cite{PLB293-94}.

The preponderance of the known solutions in five-dimensional, $\cN = 2$ supergravity resides in
the $U(1)^3$ case, specifically speaking, the $STU$ model with $n = 2$ vector multiplets. This model
has yielded an extensive range of exact solutions. Its journey began with the static non-extremal
three-charge black hole \cite{PLB383-151}. This cornerstone solution was later extended to include
rotation \cite{NPB476-118,PRD54-2612} and AdS asymptotics \cite{NPB553-317}. However, achieving
the full generalization, that is, a non-extremal, rotating charged AdS$_5$ black hole with three
independent charges and two different rotation parameters, proved to be rather difficult to
explore. Initial successes were limited to special cases: two equal rotations \cite{PRD70-081502}
or constrained charges \cite{PRD72-041901,PLB658-64}. The general solution, known as the ``Wu black
hole" \cite{PLB707-286}, was finally constructed via a novel extension of the famous Kerr-Schild
ans\"atz to supergravity theories. Subsequent studies have further dissected through investigations
of thermodynamics \cite{PLB383-151,NPB476-118,PRD54-2612,NPB553-317,PRD70-081502,PRD72-041901,
PLB658-64,PLB707-286,PLB726-404}, and symmetries \cite{1608.05052}, etc.

In contrast, the territory of $\cN = 2$ supergravity with $n > 2$ is less charted. A step beyond
the $STU$ model was taken by Giusto and Russo \cite{CQG29-085006}, who introduced a fourth charge
perturbatively in supersymmetric black rings and found a system that can be consistently truncated
to the five-dimensional theory with three vector multiplets ($n = 3$). While this extended framework
has yielded supersymmetric multi-charge solutions \cite{CQG29-085006,JHEP0512033,JHEP0712107}, the
non-extremal black hole solutions have remained elusive. These challenges have now begun to be
addressed, most notably by our recent work reporting the first construction of a static non-extremal
black hole in this framework \cite{2510.13655}. In this article, we present a new static non-extremal
AdS$_5$ black hole solution with four independent electric charge parameters in the five-dimensional
$\cN = 2$ gauged supergravity theory coupled to three vector multiplets.

On the other hand, topology is increasingly applied in black hole physics, focusing on two main
areas: (1) the analysis of light rings \cite{PRL119-251102,PRL124-181101,PRD102-064039,PRD103-104031,
PLB858-139052,PLB868-139742} and their extension to timelike orbits \cite{PRD107-064006,JCAP0723049,
PRD108-044077,CQG42-025020}, which are critical for observational astronomy, and (2) the topological
classification of black hole thermodynamics \cite{PRL129-191101,PRD110-L081501,PRD111-L061501,2509.03308,
2508.01614,PRD107-024024,PRD107-064023,PRD107-084002,EPJC83-365,EPJC83-589,PRD108-084041,JHEP0624213,
PLB856-138919,PDU46-101617,EPJC84-1294,CQG42-125007,PLB860-139163,PLB865-139482,EPJC85-828}. In this
work, after providing a comprehensive analysis of thermodynamic properties of our new static non-extremal
AdS$_5$ black hole solution, we also include a detailed investigation of the thermodynamic topological
classification of the black hole states, which reveals new topological subclasses beyond the previously
known categories.

The remaining part of this paper is organized as follows. In Sec. \ref{sec:solution}, we introduce the
five-dimensional $\cN = 2$, $STU-W^2U$ gauged supergravity theory and present our new static non-extremal
AdS$_5$ black hole solution, which is obtained by a minimal AdS extension of the four-charge
static solution we recently constructed in ungauged supergravity \cite{2510.13655}. In Sec. \ref{sec:thermo},
we display the thermodynamic quantities, showing that they satisfy the first law and Bekenstein-Smarr mass
formula, and perform a thermodynamic topological classification. Finally, we conclude in Sec.
\ref{sec:conclusion} with a summary and outlook for future work.

\section{Static non-extremal four-charge AdS black hole solution}
\label{sec:solution}

We begin by recapping the five-dimensional $\cN = 2$, $STU-W^2U$ gauged supergravity theory that
serves as the foundation for our black hole solution. The model is defined by the pre-potential
\be
\cV = \frac{1}{6}C_{IJK}X^IX^JX^K = 1 \, ,
\ee
where the non-vanishing components of the symmetric tensor $C_{IJK}$ are $C_{123} = 1$ and $C_{344}
= C_{434} = C_{443} = -2$. This pre-potential simplifies to the physically transparent form:
\be
\cV = X^1X^2X^3 -(X^4)^2X^3 = STU -W^2U \equiv 1 \, .
\ee
The reduction to the $STU$ model is immediate upon setting $X^4 = W = 0$, which physically corresponds
to switching off the fourth gauge field.

The bosonic Lagrangian of the five-dimensional $\mathcal{N}=2$ gauged supergravity coupled to three
vector multiplets is given by:
\be\begin{aligned}\label{L}
\mathcal{L} = &\frac{1}{16\pi G}\int d^{5}x\bigg\{
 \sqrt{-g}\bigg[ R -3(\partial\varphi_1)^2 -\alpha(\partial\varphi_2)^2
 -\frac{1}{4\alpha(\alpha-1)}(\partial\alpha)^2 \\
&\quad -\frac{\alpha}{4}\big(e^{-2\varphi_1-2\varphi_2}F_1^2 +e^{-2\varphi_1+2\varphi_2}F_2^2\big)
 -\frac{1}{4}e^{4\varphi_1}F_3^2 -\frac{2\alpha-1}{2}e^{-2\varphi_1}F_4^2 \\
&\quad -\frac{\alpha-1}{2}e^{-2\varphi_1}F_1F_2 +\sqrt{\alpha(\alpha-1)}
 \big(e^{-2\varphi_1-\varphi_2}F_1F_4 +e^{-2\varphi_1 +\varphi_2}F_2F_4\big)\bigg] \\
&\quad +\frac{1}{8}\varepsilon^{\mu\nu\alpha\beta\lambda}\big(F_{1\mu\nu}F_{2\alpha\beta}A_{3\lambda}
 -F_{4\mu\nu}F_{4\alpha\beta}A_{3\lambda}\big) -\frac{2}{l^2}\big[\sqrt{\alpha}\big(e^{-\phi_2-\phi_1}
 +e^{\phi_2-\phi_1}\big) +e^{2\phi_1}\big]\bigg\} \, ,
\end{aligned}
\ee
where $l$ is the cosmological scale, the field strength 2-forms $F^I = dA^I\equiv F^I_{\mu\nu}dx^{\mu}
\wedge dx^{\nu}$ correspond to the four $U(1)$ Abelian gauge field 1-forms $A^I = A^I_{\mu}dx^{\mu}$,
and the three dilaton scalar fields $(\varphi_1, \varphi_2, \alpha)$ are related to the scalars $X^I$
as follows:
\be\begin{aligned}
X^1&= \sqrt{\alpha}\, e^{\varphi_1+\varphi_2} \equiv S \, ,
& X^2&= \sqrt{\alpha}\, e^{\varphi_1-\varphi_2} \equiv T \, , \\
X^3&= e^{-2\varphi_1} \equiv U \, ,
& X^4&= \sqrt{\alpha-1}\, e^{\varphi_1} \equiv W \, .
\end{aligned}\ee

\subsection{Solution construction and parametrization}
To construct the general static non-extremal four-charge AdS$_5$ black hole, we employ an ansatz that
naturally generalizes known solutions \cite{NPB553-317} while introducing the fourth charge parameter.
This solution is based on our earlier construction of a general static non-extremal black hole in the
framework of five-dimensional $\cN = 2$ supergravity coupled to three vector multiplets \cite{2510.13655},
extended to the AdS case by including the last term in Eq. (\ref{L}). The detailed construction of the general
solution can be found in Sec. 3 of Ref. \cite{2510.13655}. Its key element is the adoption of the following
parametrization for the scalar fields:
\be
\varphi_1 = \frac{1}{6} \ln \Big(\frac{Z_3^2}{Z_1Z_2 -Z_4^2}\Big) \, , \quad
\varphi_2 = \frac{1}{2} \ln \Big(\frac{Z_2}{Z_1}\Big) \, , \quad
\alpha = \frac{Z_1Z_2}{Z_1Z_2 -Z_4^2}\, ,
\ee
where the functions $Z_I$ are defined as:
\be
Z_{I} = 1 +\frac{q_{I}}{r^2} \quad (I=1,2,3)\, , \qquad Z_4 = \frac{q_4}{r^2} \, ,
\ee
Here, $q_I$ are the electric charge parameters. The appearance of the combination $Z_1Z_2 -Z_4^2$ reflects the
distinctive structure of the $STU-W^2U$ prepotential.

\subsection{Metric and gauge potentials}
With this parametrization, the metric and gauge potentials take the form:
\be\begin{aligned}\label{metric}
ds^2 &= (Z_1Z_2 -Z_4^2)^{1/3}Z_3^{1/3}\Big[-\frac{f(r)dt^2}{(Z_1Z_2 -Z_4^2)Z_3}
 +\frac{dr^2}{f(r)} +r^2d\Omega_3^2\Big] \, , \\
A_1 &= \frac{p_1Z_2 -p_4Z_4}{r^2(Z_1Z_2 -Z_4^2)}dt \, , \qquad
A_2 = \frac{p_2Z_1 -p_4Z_4}{r^2(Z_1Z_2 -Z_4^2)}dt \, , \\
A_3 &= \frac{p_3}{r^2Z_3}dt \, , \qquad\qquad
A_4 = \frac{q_4(p_2Z_1 -p_1Z_2)}{(q_1-q_2)r^2(Z_1Z_2 -Z_4^2)}dt \, ,
\end{aligned}\ee
where $d\Omega_3^2 = d\theta^2 +\sin^2\theta d\phi^2 +\cos^2\theta d\psi^2$ is the metric on
the unit 3-sphere. The various functions and related constants are defined as follows:
\be\begin{aligned}
f(r) &= 1 -\frac{2m}{r^2} +\frac{p_4^2 +(\bar{w}-1)q_4^2}{r^4}
 +\frac{r^2}{l^2}(Z_1Z_2 -Z_4^2)Z_3 \, , \\
p_1 &= \sqrt{q_1^2 +2m q_1 +\bar{w}q_4^2} \, , \qquad
p_2 = \sqrt{q_2^2 +2m q_2 +\bar{w} q_4^2} \, , \\
p_3 &= \sqrt{q_3^2 +2m q_3 +p_4^2 +(\bar{w}-1)q_4^2} \, , \quad
p_4 = q_4\frac{p_1 -p_2}{q_1 -q_2} \, .
\label{fps}
\end{aligned}\ee
Here, $m$ is the mass parameter, $q_I$ are the electric charge parameters, $\bar{w}$ is an arbitrary
constant with two conventional choices: $\bar{w}=0$ or $\bar{w}=1$. When reduces to the STU model, parameters $s_I = \sinh\delta_I$, $c_I = \cosh\delta_I$ are often used to describe the charges. Then, we have $q_I = 2ms_I^2$, $p_I = 2ms_Ic_I$ ($I = 1,2,3$).

The consistency of the full set of field equations imposes a constraint among the parameters,
which can be expressed as:
\be
p_3^2 = p_4^2 + (\bar{w}-1)q_4^2 + (q_3-q_2)(q_3-q_1) + \frac{(q_3-q_2)p_1^2 - (q_3-q_1)p_2^2}{q_1-q_2}.
\ee
This relation, together with Eq. (\ref{fps}), completes the specification of the solution.

\subsection{Solution verification and reduction}
The form of $f(r)$ is obtained by substituting the above metric and gauge potentials into the field
equations of the $STU-W^2U$ model and solving them. The detailed derivation (see Ref. \cite{2510.13655})
shows that $f(r)$ satisfies a specific differential equation whose general solution takes the form
$f(r) = 1 +f_2/r^2 +f_4/r^4$. The specific solution given in (\ref{metric}) is the one that satisfies all
field equations and exhibits the correct asymptotic behavior.

The transition from the ungauged to the AdS case is remarkably simple--the metric function
$f(r)$ acquires an additional modification: $r^2(Z_1Z_2 -Z_4^2)Z_3/l^2$, which is proportional to the
cosmological constant, while the gauge potentials and scalar fields retain their forms from the ungauged
solution \cite{2510.13655}.

When $q_4 = p_4 = 0$, our solution reduces to the known static non-extremal three-charge AdS$_5$
black hole \cite{NPB553-317}, demonstrating the consistency of our generalization.

\section{Thermodynamics of the Four-Charge AdS Solution}
\label{sec:thermo}

\subsection{Thermodynamical properties}

Having established the geometrical structure of the solution, we now examine its thermodynamic
properties. The black hole possesses a Killing horizon at $r = r_h$, determined by the largest
positive root of $f(r_h) = 0$. The Bekenstein-Hawking entropy and Hawking temperature are given
by:
\be
S = \frac{1}{2}\pi^2 r^3 \sqrt{Z_3(Z_1Z_2 -Z_4^2)} \Big|_{r=r_h} \, , \quad
T = \frac{\p_r f(r)}{2(Z_1Z_2 -Z_4^2)Z_3} \Big|_{r=r_h} \, .
\ee
These expressions have clear physical interpretations: the entropy is proportional to the horizon
area, while the temperature reflects the surface gravity at the horizon.

The electrostatic potentials computed via $\Phi_I = A_t^{I}(r_h)$ at the horizon, conjugate to
the electric charges, are:
\be\begin{aligned}
\Phi_1 &= \frac{(p_1Z_2 -p_4Z_4)}{r^2(Z_1Z_2 -Z_4^2)}\Big|_{r=r_h} \, , \quad
&\Phi_2 &= \frac{(p_2Z_1 -p_4Z_4)}{r^2(Z_1Z_2 -Z_4^2)}\Big|_{r=r_h} \, , \\
\Phi_3 &= \frac{p_3}{r^2Z_3}\Big|_{r=r_h} \, , \quad
&\Phi_4 &= \frac{q_4(p_2Z_1 -p_1Z_2)}{(q_1-q_2)r^2(Z_1Z_2 -Z_4^2)}\Big|_{r=r_h} \, .
\end{aligned}\ee
The physical electric charges calculated via the Gauss's integral at infinity are:
\be
Q_{I} = \frac{1}{4}\pi p_I \quad (I = 1,2,3,4) \, .
\ee

Using the Abbott-Deser method \cite{NPB195-76,PRD73-104036}, the mass of the black hole is
found to be:
\be
M = \frac{1}{4}\pi (3m +q_1 +q_2 +q_3) \, ,
\ee
which represents the total energy contained in the spacetime. In the extended phase space formalism
\cite{PRD84-024037}, this mass-energy is often reinterpreted as a thermodynamic enthalpy, and the
contribution of the cosmological constant is conventionally treated as a thermodynamic pressure:
\be
P = \frac{(D-1)(D-2)}{16\pi\, l^2} = \frac{3}{4\pi\, l^2} \, ,
\ee
with its conjugate variable acted as a thermodynamic volume $V$. Remarkably, all thermodynamic
quantities satisfy both the differential and integral forms of the mass formula:
\be
\begin{aligned}
dM &= T dS +\Phi_1dQ_1 +\Phi_2dQ_2 +\Phi_3dQ_3 -2\Phi_4dQ_4 +VdP \, , \\
M &= \frac{3}{2}T S +\Phi_1Q_1 +\Phi_2Q_2 +\Phi_3Q_3 -2\Phi_4Q_4 -VP \, .
\end{aligned}
\ee
The unusual factor of `$-2$' for the work terms: $\Phi_4Q_4$ and $\Phi_4dQ_4$, stems from the
specific structure of the pre-potential $\mathcal{V} = STU -W^2U$, which leads to a non-standard
normalization for the kinetic term of the fourth gauge field $A_4$ in the Lagrangian (\ref{L}).
The thermodynamic volume, conjugate to the pressure $P$, is given by
\be
V = \frac{1}{6}\pi^2\big[3r_h^4 +2(q_1+q_2+q_3)r_h^2 +(q_1+q_2)q_3 +q_1q_2 -q_4^2\big] \, .
\ee
The satisfaction of these relations provides strong evidence for the consistency of our solution
and its thermodynamic interpretation. The inclusion of the pressure-volume work term completes
the analogy between black hole mechanics and conventional thermodynamics, offering a richer
framework for exploring the phase structure of AdS black holes.

\subsection{Thermodynamic topological classifications}

In this subsection, we shall further investigate thermodynamic topological classification of the
five-dimensional static non-extremal AdS black hole with four electric charges, utilizing the
thermodynamic quantities derived above. Let us begin with a brief review of the thermodynamic
topological method proposed in Refs. \cite{PRL129-191101,PRD110-L081501,PRD111-L061501,2509.03308}.
In this framework, the generalized off-shell Helmholtz free energy is defined as
\be\label{FE}
\mathcal{F} = M -\frac{S}{\tau} \, .
\ee
Here, $M$ and $S$ represent the mass and entropy of the black hole, respectively. The system is kept
off-shell by the parameter $\tau$, an extra variable that has the dimension of time, which can be
thought of as the inverse temperature of the cavity enclosing the black hole. It is vital to note
that the on-shell condition is met if and only if $\tau$ is equal to the inverse Hawking temperature,
i.e., $\tau = 1/T$. Under this condition, the generalized off-shell free energy simplifies to its
standard form: $F = M -TS$ \cite{PRD15-2752,PRD33-2092}.

To proceed, we incorporate an additional parameter $\Theta\in(0, \pi)$. This enables the definition
of a two-component vector
\be
\phi = \big(\phi^{r_h}, \phi^\Theta\big)
= \Big(\frac{\p\mathcal{F}}{\p\, r_h}, -\,\frac{\cos\Theta}{\sin^2\Theta}\Big) \, .
\ee
The zero points of the vector field, which correspond to black hole solutions at the inverse temperature $\tau$, are analyzed as follows: The second component vanishes only if $\Theta = \pi/2$, so the first component demands closer scrutiny. Its vanishing condition,
\be
\phi^{r_h} = \frac{\p M}{\p S}\frac{\p S}{\p r_h} -\frac{1}{\tau}\frac{\p S}{\p r_h}
= \frac{\p S}{\p r_h}\Big(T -\frac{1}{\tau}\Big) = 0 \, ,
\ee
namely, $\tau = 1/T$, thereby defines the relevant solutions.

Within Duan's $\phi$-mapping topological current formalism \cite{SS9-1072,NPB514-705},
one can construct a conserved current associated with the zeros of a two-component field $\phi^{a}$.
We work in the parameter space $x^{\nu}=(\tau,r_{h},\Theta)$ and introduce the normalized vector field $n^{a}=\phi^{a}/\|\phi\|$, whose components are $n^{r}=\phi^{r_{h}}/\|\phi\|$ and
$n^{\Theta}=\phi^{\Theta}/\|\phi\|$. The corresponding topological current is defined as \cite{SS9-1072,
NPB514-705}
\be
j^{\mu} =\frac{1}{2\pi}\epsilon^{\mu\nu\rho}\epsilon_{ab}\p_{\nu}n^{a}\partial_{\rho}n^{b},
\qquad \mu,\nu,\rho=0,1,2.
\label{eq:top_current}
\ee
By construction, this current is identically conserved, $\partial_{\mu}j^{\mu}=0$, independent of
the dynamical equations.

The current can be equivalently rewritten in terms of the Jacobian determinant $J^{\mu}(\phi/x)$
as
\be
j^{\mu} =\delta^{2}(\phi)\,J^{\mu}\!\left(\frac{\phi}{x}\right),
\label{eq:delta_current}
\ee
which makes explicit that $j^{\mu}$ has support only at isolated points $x_{i}$ where $\phi^{a}(x_{i})=0$.
The global topological number associated with a region $\Sigma$ is defined by integrating the temporal
component,
\be
W = \int_{\Sigma} j^{0}\, d^{2}x .
\label{eq:global_W}
\ee
Using the $\delta$-function representation, this quantity reduces to a discrete sum over all zero points,
\be
W = \sum_{i=1}^{N}\beta_{i}\eta_{i} = \sum_{i=1}^{N}w_{i}.
\label{eq:W_sum}
\ee
Here $\beta_{i}$ denotes the Hopf index, characterizing how many times the mapping $\phi^{a}$ wraps its
internal space as $x^{\mu}$ encircles the $i$th zero, while $\eta_{i} = \mathrm{sign}\!\left[J^{0}(\phi/x)\right]_{z_{i}}
= \pm 1$ is the Brouwer degree, encoding the local orientation of the mapping. Their product defines the
winding number $w_{i}$ associated with each isolated zero point. This integer is an intrinsic property of the
zero point and is independent of the choice of enclosing surface. In particular, the total topological number
$W$ vanishes for any region $\Sigma$ that contains no zeros of $\phi^{a}$.

Within this framework, a direct correspondence emerges between topology and stability: locally
stable black hole states carry a winding number $w = +1$, whereas unstable states are characterized by $w = -1$. Consequently, the global topological number $W$ provides a robust and model-independent classification of black
hole branches based purely on topological invariants.

For the four-charge static non-extremal AdS black hole in the five-dimensional $\cN = 2$, $STU-W^2U$
gauged supergravity, the generalized off-shell Helmholtz free energy is
\bea
\mathcal{F} &=& \frac{3\pi r_h^4}{8l^2} +\frac{3\pi r_h^2}{8l^2}(l^2 +q_1 +q_2 +q_3)
+\frac{\pi}{8l^2}\big[2l^2(q_1 +q_2 +q_3) +3(q_1q_2 +q_1q_3 +q_2q_3 -q_4^2)\big] \nn \\
&&+\frac{3\pi}{8l^2r_h^2}\big\{\big[p_4^2 +(\bar{w}-1)q_4^2 \big]l^2 +q_1q_2q_3 -q_3q_4^2 \big\}
-\frac{\pi^2}{2\tau}\sqrt{r_h^2 +q_3}\sqrt{r_h^4 +r_h^2(q_1 +q_2) +q_1q_2 -q_4^2} \, .
\eea
Then the components of the vector $\phi$ can be computed as
\bea
\phi^{r_h} &=& \frac{3\pi}{4r_h^3}\big[r_h^4 -p_4^2 -(\bar{w}-1)q_4^2 \big]
 +\frac{3\pi}{4r_h^3l^2}\big[2r_h^6 +r_h^4(q_1 +q_2 +q_3) -q_3(q_1q_2 -q_4^2)\big] \nn \\
 &&-\frac{\pi^2r_h\big[3r_h^4 +2r_h^2(q_1 +q_2 +q_3) +q_1q_2 +q_1q_3 +q_2q_3
  -q_4^2 \big]}{2\tau\sqrt{r_h^2 +q_3}\sqrt{r_h^4 +r_h^2(q_1 +q_2) +q_1q_2 -q_4^2}} \, , \\
\phi^{\Theta} &=& -\cot\Theta\csc\Theta \, .
\eea
Therefore, the zero point of the vector can be easily given as
\bea
\tau = \frac{2\pi r_h^4l^2\big[3r_h^4 +2r_h^2(q_1 +q_2 +q_3) +q_1q_2 +q_1q_3 +q_2q_3
 -q_4^2 \big]}{3\sqrt{r_h^2 +q_3}\sqrt{r_h^4 +r_h^2(q_1 +q_2) +q_1q_2 -q_4^2}
\big\{2r_h^6 +r_h^4(q_1 +q_2 +q_3 +l^2) +l^2[(1-\bar{w})q_4^2 -p_4^2] +q_3(q_4^2 -q_1q_2)\big\}} \, .
\eea

\begin{figure}[htbp]
\subfigure[]
{\label{fig1a}
\includegraphics[width=0.3\textwidth]{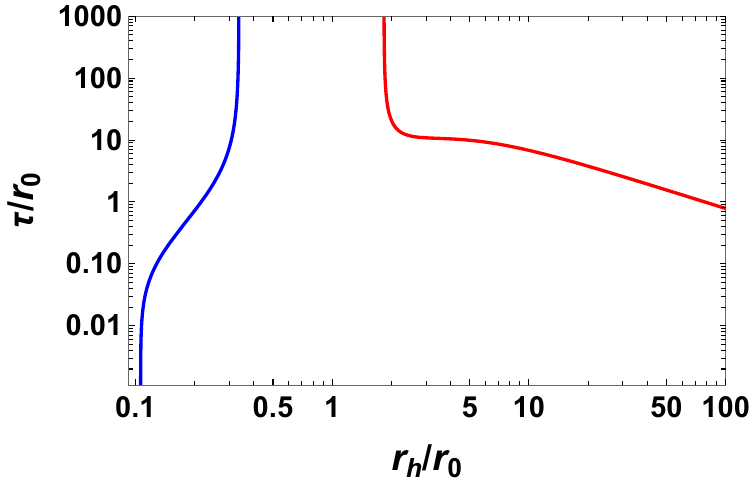}}
\subfigure[]
{\label{fig1b}
\includegraphics[width=0.3\textwidth]{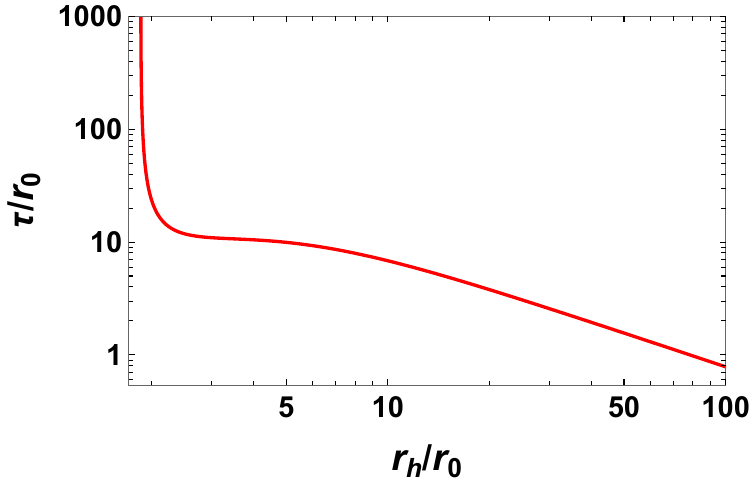}}
\subfigure[]
{\label{fig1c}
\includegraphics[width=0.3\textwidth]{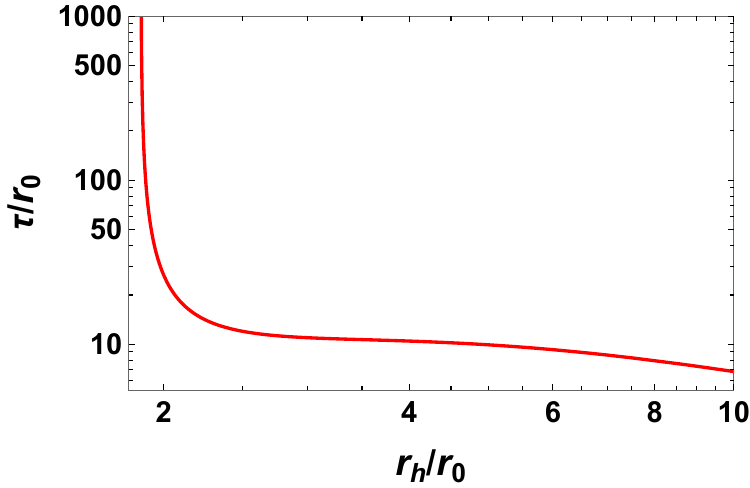}}
\caption{Zero points of $\phi^{r_h}$ in the $\tau-r_h$ plane for the four-charge static non-extremal
AdS$_5$ black hole in the $\cN = 2$, $STU-W^2U$ gauged supergravity with $q_1 = r_0$, $q_2 = 2r_0$,
$q_3 = 3r_0$, $q_4 = 0.5r_0$, $l = 5r_0$, and (a) $\bar{w} = 0$; (b) $\bar{w} = 1$; (c) $\bar{w}
= 2$, respectively. The blue line corresponds to a thermodynamically unstable black hole branch
with the winding number $w = -1$, while the red line corresponds to a thermodynamically stable
black hole branch with the winding number $w = 1$.} \label{fig1}
\end{figure}

Although $\bar{w}$ is initially an arbitrary constant in Eq. (\ref{fps}), we herein will treat it
as a continuous variable to coherently reflect the variations of the four physical electric charges
$Q_I$. As $\bar{w}$ increases, $Q_1$, $Q_2$, and $Q_3$ increase monotonically, while $Q_4$ decreases
monotonically. These results are illuminatingly reflected in Table \ref{TableI} with $\bar{w} = 0,1,2$.
With the parameters $q_1 = r_0$, $q_2 = 2r_0$, $q_3 = 3r_0$, $q_4 = 0.5r_0$, and $l = 5r_0$, Fig.
\ref{fig1} plots the zero points of $\phi^{r_h}$ in the $\tau-r_h$ plane for the constant $\bar{w} =
0, 1, 2$. Figure \ref{fig1a} $(\bar{w} = 0)$ reveals a topological number of $W = 0$. In contrast,
both Figs. \ref{fig1b} $(\bar{w} = 1)$ and \ref{fig1c} $(\bar{w} = 2)$ yield a topological number
of $W = 1$. We have also conducted a supplementary investigation by varying the AdS radius $l$ while
keeping all other parameters fixed. Our results indicate that this alteration does not affect the
thermodynamic topological classification. Therefore, a detailed discussion is omitted here.
\begin{center}
\begin{table}[htbp]
\caption{Physical charges $Q_I/r_0$ versus the parameter $\bar{w}$ for $q_1 = r_0$, $q_2 = 2r_0$,
$q_3 = 3r_0$, $q_4 = 0.5r_0$, $m = 2r_0$, and $l = 5r_0$.}\label{TableI}
\begin{tabular}{c|c|c|c|c}\hline\hline
$\bar{w}$  & $Q_1/r_0$  & $Q_2/r_0$ & $Q_3/r_0$ & $Q_4/r_0$ \\ \hline
$0$& 1.756  & 2.721 & 3.610 & 0.482 \\
$1$& 1.800  & 2.749 & 3.630 & 0.475  \\
$2$& 1.842 & 2.777 & 3.651 & 0.467  \\
\hline\hline
\end{tabular}\end{table}
\end{center}

From Fig. \ref{fig1}, one can directly read off that for $\bar{w}=0$ case, the inverse
temperature $\tau$ vanishes both at the minimal horizon radius $r_m$ and in the asymptotic limit
$r\to\infty$. In contrast, for $\bar{w} = 1$ or $2$ cases, $\tau$ diverges at $r = r_m$ while approaching
zero at infinity, where $r_m$ denotes the minimal radius of the black hole event horizon. This
behavior can be summarized as
\bea
\bar{w} = 0 &:& \tau(r_m) = 0 \, ,\qquad \tau(\infty) = 0 \, , \\
\bar{w} = 1,2 &:& \tau(r_m) = \infty \, , \qquad \tau(\infty) = 0 \, .
\eea

Next, we analyze the systematic ordering of the zero points for these three cases. In the $\bar{w} = 0$
case, there exists at least one black hole state with a negative heat capacity (corresponding to a
winding number of $-1$) and one with a positive heat capacity (corresponding to a winding number of
$+1$). The generation or annihilation of new states typically occurs in pairs with opposite
winding numbers, $(+1,-1)$ or $(-1,+1)$, ensuring the continuity of the total topological number under
smooth variations of parameters. As a result, any additional states must emerge in pairs. Since the signs
of the heat capacities alternate with increasing $r_h$, the smallest state is unstable, and the largest
is stable. Consequently, the sequence of winding numbers, ordered from the smallest to the largest $r_h$,
follows the pattern $[-, (+, -), ..., +]$, where the ellipsis denotes any number of repeating $(+,-)$ pairs.
In contrast, the behavior for the $\bar{w} = 1$ and $\bar{w} = 2$ cases differs, exhibiting the pattern $[+,
(-, +), ..., (-, +)]$. Here, both the innermost (smallest $r_h$) and outermost (largest
$r_h$) states are stable black holes.

Then, we examine the asymptotic thermodynamic behavior for the three cases. In the low-temperature
limit $(\tau \to \infty)$, the $\bar{w} = 0$ case features one unstable small and one stable small
black hole state. In contrast, only a single stable small black hole state exists for both $\bar{w}
= 1$ and $\bar{w} = 2$ cases. In the high-temperature limit $(\tau \to 0$), the behavior again
differs: the $\bar{w} = 0$ case exhibits one unstable small black hole and one stable large black
hole, while the $\bar{w} = 1$ and $\bar{w} = 2$ cases each display a single stable large black hole
state.

In conclusion, our analysis demonstrates a distinct topological classification for the four-charge
static non-extremal AdS$_5$ black hole in the $\cN = 2$, $STU-W^2U$ gauged supergravity. For the
cases $\bar{w} = 1$ and $\bar{w} = 2$, the black hole belongs to the known topological class $W^{1+}$.
In contrast, the $\bar{w} = 0$ case does not fall into any established topological category. It
instead defines a new topological subclass, which, following the standard nomenclature, we designate
as $\bar{W}^{0-}$. For convenience and clarity, we present a summary of the aforementioned
thermodynamic properties in Table \ref{TableII}.
\begin{center}
\begin{table}[htbp]
\caption{Thermodynamical properties of the black hole states for the four-charge static non-extremal
AdS$_5$ black hole in $\cN = 2$, $STU-W^2U$ gauged supergravity.}\label{TableII}
\begin{tabular}{c|c|c|c|c|c|c|c}\hline\hline
Cases & Classes  & Innermost & Outermost & Low $T$ ($\tau\to\infty$)
 & High $T$ ($\tau\to 0$) & DP & $W$ \\ \hline
$\bar{w} = 0$& $\bar{W}^{0-}$  & unstable & stable & unstable small + stable small
& unstable small + stable large & in pairs & $0$ \\
$\bar{w} = 1, 2$ & $W^{1+}$  & stable  & stable
& stable small & stable large & in pairs & $+1$ \\
\hline\hline
\end{tabular}\end{table}
\end{center}

The underlying reason for this classification is the role of the parameter $\bar{w}$ in scaling
the physical electric charges $Q_I$. For fixed $q_I$, $m$, and $l$, a change in $\bar{w}$ induces
a specific pattern of variations in $Q_I$. At small $\bar{w}$ (corresponding to smaller $Q_1$, $Q_2$,
$Q_3$ but larger $Q_4$), the system exhibits two coexisting black hole branches: one thermodynamically
stable and the other unstable. As $\bar{w}$ increases (leading to larger $Q_1$, $Q_2$, $Q_3$ and
smaller $Q_4$), the unstable branch vanishes, leaving only a single stable branch. This charge-driven
transition in the thermodynamic structure parallels that previously observed across several known
systems \cite{PRD111-L061501,JHEP0624213}: the four-dimensional static charged AdS black hole in
the Einstein-Maxwell-dilaton-axion gauged supergravity, the four-dimensional static two-charge AdS
black hole, and the five-dimensional static charged AdS black hole in Kaluza-Klein gauged supergravity.

\section{Conclusions}
\label{sec:conclusion}

In this paper, we have constructed a new static non-extremal four-charge AdS black hole solution in
the five-dimensional $\cN = 2$, $STU-W^2U$ gauged supergravity. This solution represents a natural
generalization of the three-charge static non-extremal AdS$_5$ black holes within the gauged $STU$
model, with the addition of a fourth electric charge parameter. The transition from the ungauged
case to the AdS case is remarkably simple, involving only a simple modification to the metric
function while preserving the structure of the gauge potentials and scalar fields. In addition,
we have computed the full set of thermodynamic quantities and demonstrated that they satisfy the
first law and Bekenstein-Smarr mass formula simultaneously.

Notably, our thermodynamic topological classification has uncovered a new topological subclass
$\bar{W}^{0-}$ for the case $\bar{w} = 0$, while the cases $\bar{w} = 1$ and $\bar{w} = 2$ belong
to the known class $W^{1+}$. The underlying reason for this classification is the role of $\bar{w}$
in governing the physical electric charges $Q_I$ in a special style. Our analysis shows that an
increase in $\bar{w}$ leads to an increase in $Q_1$, $Q_2$, $Q_3$ but a decrease in $Q_4$. This
specific charge configuration is what drives the transition in thermodynamic phase structure: at
smaller $\bar{w}$, two branches (one stable, one unstable) coexist, giving rise to the new subclass;
at larger $\bar{w}$, only a single stable branch remains. This observed charge-driven transition
parallels findings in other static AdS black hole systems, which significantly enriches our
understanding of the universal thermodynamic properties of multi-charge black holes in gauged
supergravity.

Looking forward, several promising directions for future research emerge. The most immediate extension
would be to incorporate rotations and/or a negative cosmological constant, leading to the most general
doubly-rotating four-charge AdS$_5$ black holes within the $STU-W^2U$ gauged supergravity model. Such
solutions would provide fertile ground for exploring the interplay between charge, rotation, and
cosmological constant in thermodynamics. This general family of exact solutions definitely provides
a new example for exploring the AdS$_5$/CFT$_4$ correspondence. On the other hand, one can also try
to seek exact charged equally-rotating solutions in the G\"odel universe, similar to the known ones
already found in Refs. \cite{PRL91-021601,PRL100-121301}.

\acknowledgments

We are grateful to the anonymous referee for useful suggestions. This work is supported
by the National Natural Science Foundation of China (NSFC) under Grants No. 12205243, No. 12375053,
and by the Sichuan Science and Technology Program under Grant No. 2026NSFSC0021.


\begin{thebibliography}{99}
\def\CQG{Classical Quantum Gravity\,}
\def\EPJC{Eur. Phys. J. C\,}
\def\JHEP{J. High Energy Phys.\,}
\def\PRD{Phys. Rev. D\,}
\def\PDU{Phys. Dark Univ.\,}
\def\PRL{Phys. Rev. Lett.\,}
\def\NPB{Nucl. Phys. B \,}
\def\PLB{Phys. Lett. B \,}

\bibitem{CQG27-205009}
D.D.K. Chow,
Symmetries of supergravity black holes,
\href{https://doi.org/10.1088/0264-9381/27/20/205009}
{\CQG \textbf{27}, 205009 (2010)}.

\bibitem{PLB293-94}
B. de Wit and A. Van Proeyen,
Broken sigma model isometries in very special geometry,
\href{https://doi.org/10.1016/0370-2693(92)91485-R}
{\PLB \textbf{293}, 94 (1992)}.

\bibitem{PLB383-151}
G.T. Horowitz, J.M. Maldacena and A. Strominger,
Nonextremal black hole microstates and U duality,
\href{https://doi.org/10.1016/0370-2693(96)00738-1}
{\PLB \textbf{383}, 151 (1996)}.

\bibitem{NPB476-118}
M. Cveti\v{c} and D. Youm,
General rotating five-dimensional black holes of toroidally compactified heterotic string,
\href{https://doi.org/10.1016/0550-3213(96)00355-0}
{\NPB \textbf{476}, 118 (1996)}.

\bibitem{PRD54-2612}
M. Cveti\v{c} and D. Youm,
Entropy of nonextreme charged rotating black holes in string theory,
\href{https://doi.org/10.1103/PhysRevD.54.2612}
{\PRD \textbf{54}, 2612 (1996)}.

\bibitem{NPB553-317}
K. Behrndt, M. Cveti\v{c} and W.A. Sabra,
Nonextreme black holes of five-dimensional $\cN = 2$ AdS supergravity,
\href{https://doi.org/10.1016/S0550-3213(99)00243-6}
{\NPB \textbf{553}, 317 (1999)}.

\bibitem{PRD70-081502}
M. Cveti\v{c}, H. L\"u and C.N. Pope,
Charged rotating black holes in five dimensional $U(1)^3$ gauged $\cN = 2$ supergravity,
\href{https://doi.org/10.1103/PhysRevD.70.081502}
{\PRD \textbf{70}, 081502 (2004)}.

\bibitem{PRD72-041901}
Z.-W. Chong, M. Cveti\v{c}, H. L\"u and C.N. Pope,
Five-dimensional gauged supergravity black holes with independent rotation parameters,
\href{https://doi.org/10.1103/PhysRevD.70.081502}
{\PRD \textbf{72}, 041901 (2005)}.

\bibitem{PLB658-64}
J.W. Mei and C.N. Pope,
New rotating non-extremal black holes in D = 5 maximal gauged supergravity,
\href{https://doi.org/10.1016/j.physletb.2007.10.045}
{\PLB \textbf{658}, 64 (2007)}.

\bibitem{PLB707-286}
S.-Q. Wu,
General nonextremal rotating charged AdS black holes in five-dimensional $U(1)^3$
gauged supergravity: A simple construction method,
\href{https://doi.org/10.1016/j.physletb.2011.12.031}
{\PLB \textbf{707}, 286 (2011)}.

\bibitem{PLB726-404}
S.-Q. Wu, D. Wen, Q.-Q. Jiang and S.-Z. Yang,
Thermodynamics of five-dimensional static three-charge STU black holes with squashed horizons,
\href{https://doi.org/10.1016/j.physletb.2013.08.019}
{\PLB \textbf{726}, 404 (2013)}.

\bibitem{1608.05052}
D.D.K. Chow,
Hidden symmetries of black holes in five-dimensional supergravity,
\href{http://arxiv.org/abs/1608.05052}{arXiv:1608.05052}.

\bibitem{CQG29-085006}
S. Giusto and R. Russo,
Adding new hair to the 3-charge black ring,
\href{https://doi.org/10.1088/0264-9381/29/8/085006}
{\CQG \textbf{29}, 085006 (2012)}.

\bibitem{JHEP0512033}
O. Vasilakis,
Bubbling the newly grown black ring hair,
\href{https://doi.org/10.1007/JHEP05(2012)033}
{\JHEP \textbf{05} (2012) 033}.

\bibitem{JHEP0712107}
B.D. Chowdhury,
Black rings with fourth dipole cause less hair loss,
\href{https://doi.org/10.1007/JHEP07(2012)107}
{\JHEP \textbf{07} (2012) 107}.

\bibitem{2510.13655}
D. Wu and S.-Q. Wu,
Four-charge static non-extremal black holes in the five-dimensional $\cN =2$, $STU-W^2U$ supergravity,
\href{http://arxiv.org/abs/2510.13655}{arXiv:2510.13655}.

\bibitem{PRL119-251102}
P.V.P. Cunha, E. Berti and C.A.R. Herdeiro,
Light Ring Stability in Ultra-Compact Objects,
\href{http://dx.doi.org/10.1103/PhysRevLett.119.251102}
{\PRL \textbf{119}, 251102 (2017)}.

\bibitem{PRL124-181101}
P.V.P. Cunha and C.A.R. Herdeiro,
Stationary Black Holes and Light Rings,
\href{http://dx.doi.org/10.1103/PhysRevLett.124.181101}
{\PRL \textbf{124}, 181101 (2020)}.

\bibitem{PRD102-064039}
S.-W. Wei,
Topological charge and black hole photon spheres,
\href{https://doi.org/10.1103/PhysRevD.102.064039}
{\PRD \textbf{102}, 0604039 (2020)}.

\bibitem{PRD103-104031}
M. Guo and S. Gao,
Universal properties of light rings for stationary axisymmetric spacetimes,
\href{https://doi.org/10.1103/PhysRevD.103.104031}
{\PRD \textbf{103}, 104031 (2021)}.

\bibitem{PLB858-139052}
W. Liu, D. Wu and J. Wang,
Light rings and shadows of static black holes in effective quantum gravity,
\href{https://doi.org/10.1016/j.physletb.2024.139052}
{\PLB \textbf{858}, 139052 (2024)}.

\bibitem{PLB868-139742}
W. Liu, D. Wu and J. Wang,
Light rings and shadows of static black holes in effective quantum gravity II: A new solution
without Cauchy horizons,
\href{https://doi.org/10.1016/j.physletb.2025.139812}
{\PLB \textbf{868}, 139742 (2025)}.

\bibitem{PRD107-064006}
S.-W. Wei and Y.-X. Liu,
Topology of equatorial timelike circular orbits around stationary black holes,
\href{https://doi.org/10.1103/PhysRevD.107.064006}
{\PRD \textbf{107}, 064006 (2023)}.

\bibitem{JCAP0723049}
X. Ye and S.-W. Wei,
Topological study of equatorial timelike circular orbit for spherically symmetric (hairy) black holes,
\href{https://doi.org/10.1088/1475-7516/2023/07/049}
{J. Cosmol. Astropart. Phys. \textbf{07} (2023) 049}.

\bibitem{PRD108-044077}
J. Yin, J. Jiang and M. Zhang,
Kinematic topologies of black holes,
\href{https://doi.org/10.1103/PhysRevD.108.044077}
{\PRD \textbf{108}, 044077 (2023)}.

\bibitem{CQG42-025020}
X. Ye and S.-W. Wei,
Novel topological phenomena of timelike circular orbits for charged test particles,
\href{https://doi.org/10.1088/1361-6382/ad9f14}
{\CQG \textbf{42}, 025020 (2025)}.

\bibitem{PRL129-191101}
S.-W. Wei, Y.-X. Liu and R.B. Mann,
Black Hole Solutions as Topological Thermodynamic Defects,
\href{https://doi.org/10.1103/PhysRevLett.129.191101}
{\PRL \textbf{129}, 191101 (2022)}.

\bibitem{PRD110-L081501}
S.-W. Wei, Y.-X. Liu and R.B. Mann,
Universal topological classifications of black hole thermodynamics,
\href{https://doi.org/10.1103/PhysRevD.110.L081501}
{\PRD \textbf{110}, L081501 (2024)}.

\bibitem{PRD111-L061501}
D. Wu, W. Liu, S.-Q. Wu and R.B. Mann,
Novel topological classes in black hole thermodynamics,
\href{https://doi.org/10.1103/PhysRevD.111.L061501}
{\PRD \textbf{111}, L061501 (2025)}.

\bibitem{2509.03308}
W. Ai and D. Wu,
$\widetilde{W}^{1+}$ subclass: Extending the topological classification of black hole thermodynamics,
\href{https://arxiv.org/abs/2509.03308}{arXiv:2509.03308}.

\bibitem{2508.01614}
S.-P. Wu, S.-J. Yang and S.-W. Wei,
Extended thermodynamical topology of black hole,
\href{https://arxiv.org/abs/2508.01614}{arXiv:2508.01614}.

\bibitem{PRD107-024024}
D. Wu,
Topological classes of rotating black holes,
\href{https://doi.org/10.1103/PhysRevD.107.024024}
{\PRD \textbf{107}, 024024 (2023)}.

\bibitem{PRD107-064023}
C.H. Liu and J. Wang,
The topological natures of the Gauss-Bonnet black hole in AdS space,
\href{https://doi.org/10.1103/PhysRevD.107.064023}
{\PRD \textbf{107}, 064023 (2023)}.

\bibitem{PRD107-084002}
D. Wu and S.-Q. Wu,
Topological classes of thermodynamics of rotating AdS black holes,
\href{https://doi.org/10.1103/PhysRevD.107.084002}
{\PRD \textbf{107}, 084002 (2023)}.

\bibitem{EPJC83-365}
D. Wu,
Classifying topology of consistent thermodynamics of the four-dimensional
neutral Lorentzian NUT-charged spacetimes,
\href{https://doi.org/10.1140/epjc/s10052-023-11561-4}
{\EPJC \textbf{83}, 365 (2023)}.

\bibitem{EPJC83-589}
D. Wu,
Consistent thermodynamics and topological classes for the four-dimensional
Lorentzian charged Taub-NUT spacetimes,
\href{https://doi.org/10.1140/epjc/s10052-023-11782-7}
{\EPJC \textbf{83}, 589 (2023)}.

\bibitem{PRD108-084041}
D. Wu,
Topological classes of thermodynamics of the four-dimensional static accelerating black holes,
\href{https://doi.org/10.1103/PhysRevD.108.084041}
{\PRD \textbf{108}, 084041 (2023)}.

\bibitem{JHEP0624213}
D. Wu, S.-Y. Gu, X.-D. Zhu, Q.-Q. Jiang and S.-Z. Yang,
Topological classes of thermodynamics of the static multi-charge AdS black holes in gauged
 supergravities: novel temperature-dependent thermodynamic topological phase transition,
\href{https://doi.org/10.1007/JHEP06(2024)213}
{\JHEP \textbf{06} (2024) 213}.

\bibitem{PLB856-138919}
X.-D. Zhu, D. Wu and D. Wen,
Topological classes of thermodynamics of the rotating charged AdS black holes in gauged supergravities,
\href{https://doi.org/10.1016/j.physletb.2024.138919}
{\PLB \textbf{856}, 138919 (2024)}.

\bibitem{PDU46-101617}
H. Chen, D. Wu, M.-Y. Zhang, H. Hassanabadi and Z.-W. Long,
Thermodynamic topology of phantom AdS black holes in massive gravity,
\href{https://doi.org/10.1016/j.dark.2024.101617}
{\PDU \textbf{46}, 101617 (2024)}.

\bibitem{EPJC84-1294}
Z.-Q. Chen and S.-W. Wei,
Thermodynamical topology with multiple defect curves for dyonic AdS black holes,
\href{https://doi.org/10.1140/epjc/s10052-024-13620-w}
{\EPJC \textbf{84}, 1294 (2024)}.

\bibitem{CQG42-125007}
W. Liu, L. Zhang, D. Wu and J. Wang,
Thermodynamic topological classes of the rotating, accelerating black holes,
\href{https://doi.org/10.1088/1361-6382/ade35b}
{\CQG \textbf{42}, 125007 (2025)}.

\bibitem{PLB860-139163}
X.-D. Zhu, W. Liu and D. Wu,
Universal thermodynamic topological classes of rotating black holes,
\href{https://doi.org/10.1016/j.physletb.2024.139163}
{\PLB \textbf{860}, 139163 (2025)}.

\bibitem{PLB865-139482}
Y. Chen, X.-D. Zhu and D. Wu,
Universal thermodynamic topological classes of three-dimensional BTZ black holes,
\href{https://doi.org/10.1016/j.physletb.2025.139482}
{\PLB \textbf{865}, 139482 (2025)}.

\bibitem{EPJC85-828}
H. Chen, D. Wu, M.-Y. Zhang, S. Zare, H. Hassanabadi, B.C. L\"utf\"uo\v{g}lu and Z.-W. Long,
Universal thermodynamic topological classes of static black holes in conformal Killing gravity,
\href{https://doi.org/10.1140/epjc/s10052-025-14581-4}
{\EPJC \textbf{85}, 828 (2025)}.

\bibitem{NPB195-76}
L.F. Abbott and S. Deser,
Stability of gravity with a cosmological constant,
\href{https://doi.org/10.1016/0550-3213(82)90049-9}
{\NPB \textbf{195}, 76 (1982)}.

\bibitem{PRD73-104036}
W. Chen, H. L\"u and C.N. Pope,
Mass of rotating black holes in gauged supergravities,
\href{https://doi.org/10.1103/PhysRevD.73.104036}
{\PRD \textbf{73}, 104036 (2006)}.

\bibitem{PRD84-024037}
M. Cveti\v{c}, G.W. Gibbons, D. Kubiz\v{n}\'ak and C.N. Pope,
Black hole enthalpy and an entropy inequality for the thermodynamic volume,
\href{https://doi.org/10.1103/PhysRevD.84.024037}
{\PRD \textbf{84}, 024037 (2011)}.

\bibitem{PRD15-2752}
G.W. Gibbons and S.W. Hawking,
Action integrals and partition functions in quantum gravity,
\href{https://doi.org/10.1103/PhysRevD.15.2752}
{\PRD \textbf{15}, 2752 (1977)}.

\bibitem{PRD33-2092}
J.W. York,
Black-hole thermodynamics and the Euclidean Einstein action,
\href{https://doi.org/10.1103/PhysRevD.33.2092}
{\PRD \textbf{33}, 2092 (1986)}.

\bibitem{SS9-1072}
Y.-S. Duan and M.-L. Ge,
SU(2) gauge theory and electrodynamics of $N$ moving magnetic monopoles,
\href{https://doi.org/10.1142/9789813237278_0001}
{Sci. Sin. \textbf{9}, 1072 (1979)}.

\bibitem{NPB514-705}
Y.-S. Duan, S. Li and G.-H. Yang,
The bifurcation theory of the Gauss-Bonnet-Chern topological current and Morse function,
\href{https://doi.org/10.1016/S0550-3213(97)00777-3}
{\NPB \textbf{514}, 705 (1998)}.

\bibitem{PRL91-021601}
E.G. Gimon and A. Hashimoto,
Black holes in G\"odel universes and pp waves,
\href{https://doi.org/10.1103/PhysRevLett.91.021601}
{\PRL \textbf{91}, 021601 (2003)}.

\bibitem{PRL100-121301}
S.-Q. Wu,
General Non-extremal Rotating Charged G\"odel Black Holes in Minimal Five-Dimensional
 Gauged Supergravity,
\href{https://doi.org/10.1103/PhysRevLett.100.121301}
{\PRL \textbf{100}, 121301 (2008)}.

\end{thebibliography}
\end{document}